# Direct numerical simulation of thermo-hydro-mechanical processes at Soultz-sous-Forêts

Saeed Mahmoodpour[1], Mrityunjay Singh[2], Ramin Mahyapour[3], Sri Kalyan Tangirala[4], Kristian Bär[5], Ingo Sass[2,6]

[1] Group of Geothermal Technologies, Technical University Munich, Munich, Germany

[2] Technische Universität Darmstadt, Institute of Applied Geosciences, Group of Geothermal Science and Technology, Schnittspahnstrasse 9, 64287 Darmstadt, Germany

[3] Sharif University of Technology, Tehran, Iran

[4] Department of Applied Geophysics, Indian Institute of Technology (Indian School of Mines) Dhanbad, India; tangiralask@gmail.com

[5] GeoThermal Engineering GmbH, Karlsruhe, Germany; baer@geo-t.de

[6] Darmstadt Graduate School of Excellence Energy Science and Engineering, Otto-Berndt-Strasse 3, 64287 Darmstadt, Germany

***Correspondence to*: Saeed Mahmoodpour (saeed.mahmoodpour@tum.de); Mrityunjay Singh (mrityunjay.singh@tu-darmstadt.de)**

**Abstract**

Porosity and permeability alteration due to the thermo-poro-elastic stress field disturbance from the cold fluid injection is a deciding factor for longer, economic and safer heat extraction from an enhanced geothermal system (EGS). In the Soultz-sous-Forêts geothermal system, faulted zones are the main flow paths and the resulting porosity-permeability development over time due to stress reorientation is more sensitive in comparison with the regions without faulted zones. Available operational and field data are combined through a validated numerical simulation model to examine the mechanical impact on the pressure and temperature evolution. Results shows that near the injection wellbore zones, permeability and porosity values are strongly affected by stress field changes and permeability changes will affect the overall temperature and pressure of the system demonstrating a fully coupled phenomenon. In some regions inside the faulted zones and close to injection wellbores, porosity doubles whereas permeability may enhance up to 30 times. From the sensitivity analysis on the two unknown parameters from the mechanical aspect is performed and results shows that only one of the parameters impacts significantly on the porosity-permeability changes. Further experimental and field works on this parameter will help to model the heat extraction more precisely than before.

# 1. Introduction

Fractured geothermal systems, also known as Enhanced Geothermal Systems (EGS) are a subset of geothermal systems where the rocks (mostly granites) have very low permeability and little or no water content. Naturally occurring fractures or fracture clusters (or even faults) might be present based the site's stress field and tectonic history. It is preferable to drill wells close to or intersecting these structures to utilize them without the need of spending time and resources to create an artificial fracture network. However, natural fractures generally do not have the kind of permeabilities we need for fluid flow. In such cases, we stimulate the fractures through hydraulic, chemical or thermal stimulation. A series of hydraulic stimulations were performed at Soultz-sous-Forêts from 1993 to 1996 at a depth of 2800-3500 m. This led to the creation of a large fracture network across an area of more than 3 km² (Baumgärtner et. al 1998). Many circulation tests were also performed at this site before the start of its commercial production in 2016. The amount of data from this site is way beyond any other HDR geothermal site in the world. We used the site data to build a comprehensive Thermo-Hydro-Mechanical (THM) model of the granite reservoir to examine the mechanical parameters impact on the pressure and temperature development during the cold fluid injection.

The Hot Dry Rock (HDR) project at Soultz-sous-Forêts, France started in 1984, for which drilling started in 1987 (Gérard et al. 2006). It is one of the first projects of its kind in mainland Europe. The term 'Enhanced Geothermal System (EGS)' has been coined at this site. Geothermal work has been going on at this site for the past thirty years which was subject to extensive geoscientific studies. The main objective of this project was to generate electricity by tapping the crystalline (granitic) section of the reservoir. Several wells were drilled over this period and commercial production started in 2016. For the commercial production, a three well system with two injector wells (GPK-3 and GPK-4) and one producer well (GPK-2) are in operation (see Figure 1). The brine is recovered from the reservoir at a temperature of 160 °C and is passed through and Organic Rankine Cycle (ORC) where the heat from thr brine is used to heat isobutane which then powers the turbine (Genter et. al 2009). After this, the cooled brine is re-injected back into the reservoir through an injector at 70 °C. Figure 1 is a schematic diagram of the wells and the parts of the reservoir the wells go through. During the MEET project, the possibility of the using colder fluid injection (40 °C) is examined to increase heat extraction amount from this site.

Soultz-sous-Forêts is situated in the Upper Rhine Graben (URG), which is a continental rift structure and extends up to a length of 300 kms making it the central section of the European Cenozoic Rift system. The stratigraphy of this site begins with Mesozoic and Cenozoic sediments on the top reaching to depths of 1.4 kms followed by the crystalline basement divided by naturally fractured granite. The sedimentary section of the site can broadly be divided into two distinct kinds- the fluvial deposits of 350 m in thickness, known as Buntsandstein and the alluvial deposits (Permian) from the Variscan orogeny (Duringer et al. 2019). It has been found that the sedimentary section has a geothermal gradient of >100 °C/km, whereas the granitic section has a gradient of 10-12 °C/km. This might be due to the drastic drop in the heat production values of granite with depth 6 $\mu W/m^3$ at a depth of 1400 m to 2.7 $\mu W/m^3$ at a depth of 3700 m (Guillou-Frottier et. al 2013).

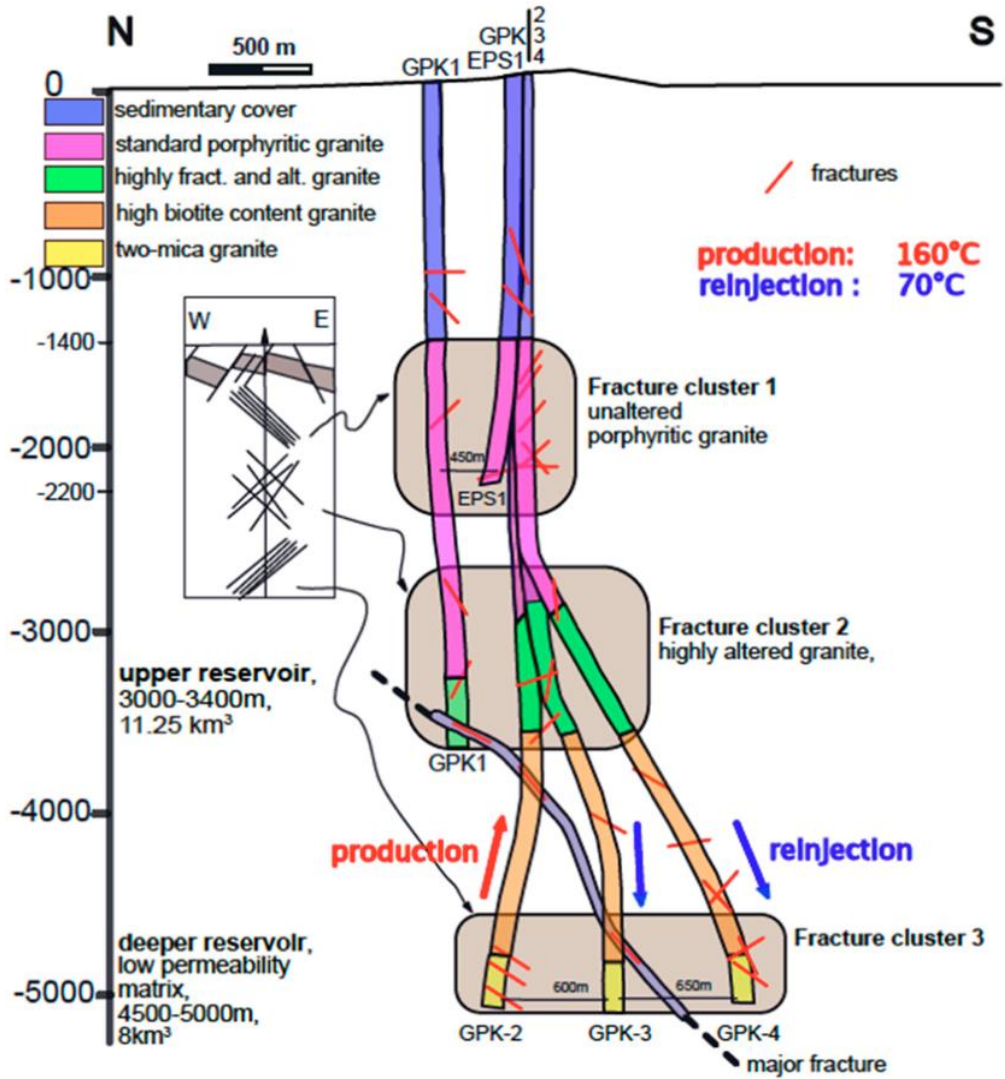

Figure 1. Schematic diagram of wells with different fracture clusters in color coded strata (Ledésert et al. 2020).

At the Soultz-sous-Forêts site, 53 structures that includes 39 fracture zones, 7 microseismic structures and 6 vertical seismic profiling (VSP) are reported by Sausse et al. (2010). Furthermore, Dezayes et al. (2010) observed that 39 fracture zones have general strike of N160°E at this geothermal site. However, in this study, only five zones of fractures are considered which have their intersection with at least one of the wellbores, and their properties are laid out in Table 1 (Boujard et al. 2021, Mahmoodpour et al. 2021a). It is to be noted that the fracture zones are numbered based on their depths of the intersection with one of the wellbores.

Table 1. Properties of the five sets of fractures (Rolin et al. 2018, Boujard et al. 2021).

| Parameter | Unit | FZ1800 | FZ2120 | FZ4760 | FZ4770 | FZ4925 |
| --- | --- | --- | --- | --- | --- | --- |

| Hydraulic conductivity | m·s$^{-1}$ | 6.08×10$^{-6}$ | 1.7×10$^{-5}$ | 0.05 | 2×10$^{-5}$ | 6.3×10$^{-5}$ |
|---|---|---|---|---|---|---|
| Specific storage | 1·m$^{-1}$ | 2×10$^{-6}$ | 2×10$^{-6}$ | 2×10$^{-6}$ | 2×10$^{-6}$ | 2×10$^{-6}$ |
| Porosity | - | 0.1 | 0.1 | 0.1 | 0.1 | 0.1 |
| Thermal conductivity | W·m$^{-1}$·K$^{-1}$ | 2.5 | 2.5 | 2.5 | 2.5 | 2.5 |
| Thermal capacity | J·m$^{-3}$K$^{-1}$ | 2.9×10$^{-6}$ | 2.9×10$^{-6}$ | 2.9×10$^{-6}$ | 2.9×10$^{-6}$ | 2.9×10$^{-6}$ |
| Thickness | m | 12 | 15 | 8 | 15 | 1 |
| Heat production | W·m$^{-3}$ | 3×10$^{6}$ | 3×10$^{6}$ | 3×10$^{6}$ | 3×10$^{6}$ | 3×10$^{6}$ |
| Transmissivity | m$^{2}$·s$^{-1}$ | 7.3×10$^{-5}$ | 2.55×10$^{-4}$ | 0.4 | 3×10$^{-4}$ | 6.3×10$^{-5}$ |

Various studies have been performed on numerical modelling of this site as well. Vallier et. al 2020 performed a Thermo-Hydro-Mechanical (THM) study which showed convection loops of temperature distribution in the reservoir of almost 1.3 kms in size. They compared these results for different permeabilities and the circulation loops appear only at a permeability of 10$^{-14}$ m$^2$. For permeabilities less than the order of -14, the loops start to diminish and are completely gone when we reach the order of -17 in the layers which accommodate circulation of fluids. Our previous Thermo-Hydro (TH) model was validated using the data of 1163 days of operation. We modelled cases with 4 different re-injection temperatures for a duration of 100 years and we observed that the thermal breakdown at the production well was less than 20 °C even after 100 years of operation (Mahmoodpour et al. 2022). This showed that the production can continue for a long time without worrying about the thermal breakdown.

Considering the available literature on the THM behavior of the heat extraction from the geothermal reservoirs, it is clear that most of the researchers used the Barton - Bandis model to update the stress field based on the aperture variations. During these studies, variation in the porosity of the fracture and the porosity – permeability variation in the matrix zone is not considered. While in most cases, due to the difficulties of considering all the fractures, equivalent porous media is assumed instead of the fractured medium. To account for all of the possible variations, the porosity and permeability variations should be considered not only for the large-scale fracture but also for the equivalent porous medium. To do this, Rutqvist et al. (2002) formulated an equation for stress-porosity dependency. This dependency may be used in the equations between the porosity and permeability. Davis and Davis (2001) provided the relationships between the porosity and permeability for different rock types which may be used to update the permeability values based on the porosity obtained from the equations developed by Rutqvist et al. (2002). In similar studies, researchers tried to use THM simulations to include the fracture aperture (or permeability) variation on the heat extraction performance of the geothermal reservoirs. Zhao et al. (2015) simulated the heat extraction from an idealized fractured hot dry rock and observed that the fracture aperture tripled for 150 °C decreases in the temperature (Zhao et al. 2015). Further, Wang et al. (2016) developed a semi-analytical correlation for the THM behavior and showed that fracture permeability may increase up to 7 times of the initial value. They also concluded that the fracture permeability is more sensitive to the cold fluid injection compared with the rock matrix. Later, Pandey and Vishal examined the sensitivity of the affecting parameters on the performance of a single-fractured geothermal system. Their study revealed that fracture aperture is controlled by the poroelasticity at the initial time and at the later times, the thermoelasticity is the main controlling factor. In this study, fracture aperture enhanced almost twice of its initial value (Pandey and Vishal 2017). In another study, Pandey et al. (2017) examined the fracture aperture alteration for different joint stiffness, thermal expansion coefficients and rock matrix permeability in a single fractured system. Here, they reported up to three times of the fracture aperture enhancement after 30 years of the operation (Pandey et al. 2017). Yao et al. (2018) used the local thermal non-equilibrium theory through the THM approach to examine the behavior of an ideal 3D-EGS system. After 80 years of the cold fluid injection, they observed up to 4 times increases in the fracture permeability near the injection wellbore. Salimzadeh et al. (2018) reported the fracture aperture changes in a single fractured system. For the cases with higher initial temperature and stress, they observed a higher change in the fracture aperture. For the highest initial temperature and stress ($\boldsymbol{T_i}$ = 250 °C and $\boldsymbol{\sigma_i}$ = 75 MPa) fracture aperture increases up to 8 times, while for the case with ($\boldsymbol{T_i}$ = 80 °C and $\boldsymbol{\sigma_i}$ = 60 MPa), fracture aperture does not even double. In other study, for the multiple-fracture system, Vik et al. (2018) showed that stress flied resulting from a fracture affects the aperture of the other fracture. In these conditions, the fracture aperture changes will be different than the single fracture model. Yuan et al. (2020) examined the THM behavior of enhanced geothermal system in the Raft River geothermal field using numerical simulations. Their simulation resulted up to 5 times in permeability enhancement in the different directions due to the cold fluid injection. Cui et al. (2021) examined a penny shaped fracture in EGS for 30 years of the operation. Due to the thermo-poro-elastic effects, fracture aperture increases up to six times of the initial value. Aliyu and Archer (2021) considered a multi-lateral fracture to examine the heat extraction from hot dry rock geothermal systems. Their simulation results show up to 10 times increases in the fracture permeability for the different spacing in the multi-lateral fracture system after 30 years of operation. In a recent study, Kang et al. (2022) reported that fracture aperture changes due to the thermoelasticity is around 22% resulting from poroelasticity after one year of the numerical simulation. However, the contribution of the thermal stress on the fracture aperture increases with time and after 30 years, its impact increases to 161% in comparison to the hydraulic effects.

To examine the possible impact of the mechanical changes on the THM processes of the Soultz-sous-Forêts, based on the geological geometry, well trajectory and the fracture zones, numerical models are developed in this study. However, geological settings and thermo-hydraulically related parameters of the Soultz-sous-Forêts are well characterized, but regarding the mechanical parameters and their impacts, there is ongoing ambiguity. To shed light on the mechanical impact by pressure and temperature evolution, the overall process is considered by assuming a high range of the changes on the mechanical parameters which couples the porosity, permeability and stress field. In the present study, we examine the proposed relationships for the porosity-permeability and the stress from the literature, and observe that it is possible to simplify to insert the mechanical impact with two parameters measured experimentally. In spite of the most previous studies, porosity and permeability changes are not only examined for the fractured zones but its impact on the matrix zone is also considered during this study. This manuscript is organized in the following manner: first we discussed about the operational condition and numerical simulation models in the methodology section followed by results and discussion with consideration of the sensitivity analysis on the mechanical parameters.

## 2. Methodology

A cubic space of each side 8 km is considered around the GPK-2, GPK-3 and GPK-4 as shown in Figure 3. Previously, we assumed homogeneous and isotropic permeability field for the matrix zone and our TH simulations resulted in a good match with the operational data for the short period. In the long period, pressure and temperature fronts are well developed which may result to the thermo-poroelasticity changes at the reservoir which is the main concern of this study. The injection flow rates for GPK-3 and GPK-4 are 19.6 kg/s and 9.7 kg/s and the production flow rate of the GPK-2 is set as 29.3 kg/s based on the recent operational data for this site (Mahmoodpour et al. 2022). As discussed previously, the five fracture zones are considered as a large fracture with an equivalent hydraulic conductivity which is obtained through the operational data (Rolin et al. 2018; Table 1). The remaining fracture zones are not considered during the numerical simulation. Previously, we showed that this simplification does not affect the overall TH behavior of the system through the matching of the numerical simulation results with the operational data. The fracture zones inside the geometry are considered as internal boundaries. The diameter of the wellbores is small, and for the simplification and making the meshing feasible, they are represented by a line. The trajectory of the wellbores is taken from the field measurements (Rolin et al. 2018).

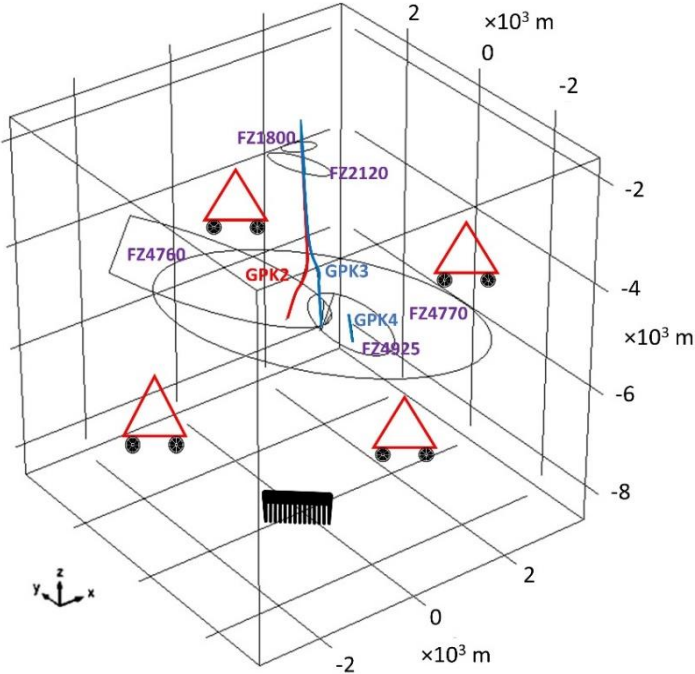

Figure 3: Geometry of the Soultz-sous-Forêts geothermal site used for numerical modeling in this study. Here the faults are numbered based on their intersection depths with the wellbores, roller boundary condition assigned at the side walls (shown by roller symbol), bottom boundary is restricted to the displacement, top boundary is free to the movement. Production well is shown with the red color while the injection wells are shown by blue colors.

We assume that mechanical changes are elastic, therefore we couple thermo-poroelasticity response of the rock to the pressure and temperature changes to update the porosity and permeability values. For the fluid flow, top and boundaries are considered as no flow boundary condition and the side boundaries are open for fluid flow. The top boundary is insulated for the heat transfer whereas a constant heat flux of 0.07 W/m² is assigned at the bottom boundary (Kohl et al. 2000). Based on the open boundary for the fluid flow, it is assumed that side boundaries are open for the heat transfer. For the geomechanical aspect, the bottom boundary is assigned with a fixed constraint boundary condition and the top boundary is assigned for free displacement. Roller boundary conditions are assigned on the sides of the numerical model geometry. The initial temperature of the system is assigned based on the geothermal gradient in the region (Pribnow and Schellschmidt 2000). Hydrostatic pressure is used for the initial pore pressure of the system. Based on operator's experience, in addition to the initial open-hole section, it is assumed that GPK-3 is injecting in the interval 1282m – 4852 m (depth reference ground level) and GPK-2 is producing from 1264 – 4244 m (Baujard et al, 2021). A homogeneous flow is assigned alongside these intervals. Here, we examined the reservoir response toward the cold fluid injection and hot water production and we did not investigate the heat loss resulting from the fluid movement through the wellbore. For more details on the wellbore effect, readers are suggested to consider Mahmoodpour et al. (2021a). The possible geochemical reaction and their impact on the porosity and permeability changes is not considered in this study. The governing equations describing the behavior of the system are explained below:

2.1 **Mass conservation**

Following governing equations are used for mass transfer, heat transfer and the stress field calculations. All of the equations are adopted from the COMSOL Multiphysics (COMSOL v5.5).

The conservation of mass transport by considering the pore-pressure, temperature and volume changes are shown by the following equation:

$$\rho_l(\phi_m S_l + (1-\phi_m)S_m)\frac{\partial p}{\partial t} - \rho_l(\alpha_m(\phi_m \beta_l + (1-\phi_m)\beta_m))\frac{\partial T}{\partial t} + \rho_l \alpha_m \frac{\partial \varepsilon_V}{\partial t} = \nabla \cdot (\frac{\rho_l k_m}{\mu}\nabla p)$$

[1]

where $\rho_l$, $\phi_m$, $S_l, S_m$, $p$, $\alpha_m$, $\beta_l, \beta_m$, $T$, $\varepsilon_V$, $k_m$, and $\mu$ are the fluid density, reservoir porosity, storage coefficient of the matrix, storage coefficient of fluid, pressure, Biot's coefficient of the matrix, thermal expansion coefficient of the rock, thermal expansion coefficient of the fluid, temperature, pore volumetric strain in reservoir, matrix permeability and the fluid viscosity, respectively. Here, $\phi_m S_l + (1-\phi_m)S_m$ and $\alpha_m(\phi_m \beta_l + (1-\phi_m)\beta_m)$ represents the storage coefficient of saturated porous media and equivalent thermal expansion coefficient of porous media.

For the fractures, the flow along the width is ignored due to the higher differences between the fracture length and fracture width. Based on this assumption, we have

$$\rho_l(\phi_f S_l + (1-\phi_f)S_f)b_h \frac{\partial p}{\partial t} - \rho_l(\alpha_f(\phi_f \beta_l + (1-\phi_f)\beta_f))b_h \frac{\partial T}{\partial t} + \rho_l \alpha_f b_h \frac{\partial \varepsilon_V}{\partial t} = \nabla_T \cdot \left(\frac{b_h \rho_l k_f}{\mu} \nabla_T p\right) + n \cdot \left(-\frac{\rho k_m}{\mu \nabla p}\right) \qquad [2]$$

where $\phi_f$, $S_f$, $\alpha_f$, $\beta_f$, $k_f$, and $b_h$ are the fracture porosity, storage coefficient of the fracture, Biot's coefficient of fracture, thermal expansion coefficient of the fracture zone, fracture permeability and the hydraulic aperture. Here, $\phi_f S_l + (1-\phi_f)S_f$ and $\alpha_f(\phi_f \beta_l + (1-\phi_f)\beta_f)$ represents the storage coefficient of saturated fracture zone and equivalent thermal expansion coefficient of fracture zone. Here, $n \cdot (-\frac{\rho k_m}{\mu \nabla p})$ is the mass flux exchange between saturated reservoir and the fracture zone.

2.2 **Energy balance**

With the line assumption for the wellbore geometry, effects of the wellbore on the simulation can be shown with a source and sink term. The latent heat of water is used to measure this term based on the temperature difference between the injection water and the local temperature of the rock. To simulate the heat transfer between the rock matrix and the fluid, the local thermal non-equilibrium model is used. Combining the heat conduction and heat exchange between the fluids for the rock matrix resulting to the governing equation.

$$(1 - \phi_m)\rho_m C_{p,m} \frac{\partial T_m}{\partial t} = \nabla \cdot \left((1 - \phi_m)\lambda_m \nabla T_m\right) + q_{ml}(T_l - T_m) \quad [3]$$

where $C_{p,m}$, $T_m$, $\lambda_m$, $q_{ml}$ and $T_l$ are the specific heat capacity of the rock matrix, rock temperature, heat conductivity of the rock matrix, rock matrix-pore fluid interface heat transfer coefficient, and pore fluid temperature, respectively.

To make a connection between the rock and the fracture for the heat transfer, it is assumed that total thermal energy leaving the rock matrix through the rock interface is received by the adjunct fracture using the following equation:

$$(1 - \phi_f)b_h\rho_f C_{p,f} \frac{\partial T_m}{\partial t} = \nabla_T \cdot \left((1 - \phi_f)b_h\lambda_f \nabla_T T_m\right) + b_h q_{fl}(T_l - T_m) + n \cdot (-(1 - \phi_m)\lambda_m \nabla T_m)$$

$$[4]$$

where $C_{p,f}$, $\lambda_f$ and $q_{fl}$ are the specific heat capacity of the fracture, heat conductivity of the fracture, and the rock fracture-fluid interface heat transfer coefficient. Here $n \cdot (-(1 - \phi_m)\lambda_m \nabla T_m)$ represents the heat flux exchange across the matrix and fracture.

The heat convection term in the energy balance equation is calculated as:

$$\phi_m \rho_l C_{p,l} \frac{\partial T_l}{\partial t} + \phi_m \rho_l C_{p,l} \left(-\frac{k_m \nabla p}{\mu}\right) \cdot \nabla T_l = \nabla \cdot (\phi_m \lambda_l \nabla T_l) + q_{ml}(T_m - T_l) \quad [5]$$

The coupled heat exchange between the fluid and fracture is:

$$\phi_f b_h \rho_l C_{p,l} \frac{\partial T_l}{\partial t} + \phi_f b_h \rho_l C_{p,l} \left(-\frac{k_f \nabla p}{\mu}\right) \cdot \nabla_T T_l = \nabla_T \cdot (\phi_f b_h \lambda_l \nabla_T T_l) + b_h q_{fl}(T_m - T_l) + n \cdot (-\phi_l \lambda_l \nabla T_l)$$

$$[6]$$

2.3 **Stress**

As it is mentioned, we assumed that the deformation is elastic. Therefore, stress-strain relationship is obtained by combining the following thermo-poroelastic equations:

$$\sigma_{ij} = 2G\varepsilon_{ij} + \lambda tr\varepsilon\delta_{ij} - \alpha_p p\delta_{ij} - \frac{2G(1+\nu)}{3(1-2\nu)}\phi_l\beta_l + (1-\phi_m)\beta_m T\delta_{ij} \qquad [7]$$

where $\sigma_{ij}, G, \lambda, tr, \varepsilon, \delta_{ij}, \alpha_p$ and $\nu$ are the total stress, Lame's 1st and 2nd constants, the trace operator, strain, Kronecker delta function, Biot's coefficient of the porous media and Poisson ratio, respectively. In this manuscript, positive value of stress is considered in tension mode.

The deformation equation of porous media is obtained by combining the equilibrium equations:

$$Gu_{i,jj} + (G+\lambda)u_{j,ji} - \alpha_p p_{,i} - \frac{2G(1+\nu)}{3(1-2\nu)}\beta_T T_{,i} + f_i = 0 \qquad [8]$$

where $u_{j,ji}$ and $f_i$ are displacement and the external body force, respectively.

During the heat extraction from the geothermal reservoirs, the rock will experience intense changes in the pressure and temperature. Pressure increases inside the fracture and temperature reduction will ultimately decrease the compressive normal stress on the fracture surfaces and the fracture aperture will increase. In a similar manner, the porosity and permeability of the matrix zone will enhance between the wells.

The adopted version of the equations for porosity changes with respect to the pressure and temperature variation is used to update the porosity values during the simulations (Rutqvist et al. 2002). In this system, fractures are the main flow pathways and their porosity is a strong function of the stress changes.

$$\frac{\varphi}{\varphi_0} = \frac{a + exp(c\sigma')}{a + exp(c\sigma'_0)} \qquad [9]$$

where $\varphi, \varphi_0, a, c, \sigma'$, and $\sigma'_0$ are the porosity of the porous medium, initial porosity of the porous medium, the ratio of residual porosity to the initial porosity, constant for the porosity-stress function, stress value and initial stress, respectively. This system is a fractured model and the ratio of the residual to the initial porosity is a small number compared to a nonfractured reservoir. Here, we performed a sensitivity analysis and considered three values of $a = 0.3, 0.5$ and $0.7$ to examine the overall process.

Rutqvist et al. (2002) used the value close to the 1.5 for $c$. Here, to examine the impact of this parameter, three values 1, 1.5 and 2 are considered for the different porosity-stress dependency. Davis and Davis (2001) proposed the relationship to correlate the porosity and permeability changes. In this study, rocks of mean pore diameter as 60 µm are considered for which the values of $\gamma$ varies between 2.68 and 3.15. Here, in this study, we assumed the value of $\gamma=3$.

$$k = k_0 e^{\gamma(\varphi/\varphi_0 - 1)} \tag{10}$$

Initial stress field are assigned based on the data reported by Valley and Evans (2007).

$$\sigma_h = 1.78\ MPa - 14.06z\ \text{MPa/km} \tag{11}$$

$$\sigma_H = 1.17\ MPa + 22.95z\ \text{MPa/km} \tag{12}$$

$$\sigma_v = 1.30\ MPa - 25.50z\ \text{MPa/km} \tag{13}$$

2.4 **Fluid properties**

The thermophysical properties of water such as dynamic viscosity, specific heat capacity, density, and thermal conductivity are updated based on the calculated based on the temperature distribution during the simulation using the following equations (COMSOL v5.5):

Dynamic viscosity:

$$\mu = 1.38 - 2.12 \times 10^{-2} \times (T - 273.15)^1 + 1.36 \times 10^{-4} \times (T - 273.15)^2 - 4.65 \times 10^{-7} \times (T - 273.15)^3 + 8.90 \times 10^{-10} \times (T - 273.15)^4 - 9.08 \times 10^{-13} \times (T - 273.15)^5 + 3.85 \times 10^{-16} \times (T - 273.15)^6 \quad (0 - 140\ °C) \tag{14}$$

$$\mu = 4.01 \times 10^{-3} - 2.11 \times 10^{-5} \times (T - 273.15)^1 + 3.86 \times 10^{-8} \times (T - 273.15)^2 - 2.40 \times 10^{-11} \times (T - 273.15)^3\ (140 - 280\ °C) \tag{15}$$

Specific heat capacity:

$$C_p = 1.20 \times 10^4 - 8.04 \times 10^1 \times (T - 273.15)^1 + 3.10 \times 10^{-1} \times (T - 273.15)^2 -$$
$$5.38 \times 10^{-4} \times (T - 273.15)^3 + 3.63 \times 10^{-7} \times (T - 273.15)^4 \qquad [16]$$

Density:

$$\rho = 1.03 \times 10^{-5} \times (T - 273.15)^3 - 1.34 \times 10^{-2} \times (T - 273.15)^2 + 4.97 \times (T - 273.15) +$$
$$4.32 \times 10^2 \qquad [17]$$

Thermal conductivity:

$$\kappa = -8.69 \times 10^{-1} + 8.95 \times 10^{-3} \times (T - 273.15)^1 - 1.58 \times 10^{-5} \times (T - 273.15)^2 +$$
$$7.98 \times 10^{-9} \times (T - 273.15)^3 \qquad [18]$$

To solve these equations in a fully coupled manner, COMSOL Multiphysics version 5.5 is used (COMSOL Multiphysics) through the finite element method. The provided methodology is previously validated with the operational data and give a good match in comparison to the benchmark problems from the literature (Bai, 2005, Mahmoodpour et al. 2022b). Details of the numerical simulations are available in Mahmoodpour et al. (2022a). Free tetrahedral meshes are used to simulate the THM process. To capture the behavior inside the fracture zone, smaller meshes are used. The total number of domain elements, boundary elements and edge elements are 61069, 4455 and 517 respectively. Further, the maximum and minimum element size for the matrix zone are 573 m and 71.6 m, respectively whereas for the fracture zone these values are 251 m and 10.7 m, respectively.

3. **Results and discussions**

Based on Eq. [9], parameters $a$ and c needs to be defined for the numerical simulation. Here, $a$ shows the ratio of the remaining porosity to the initial porosity in the known stress field. Due to the involved fractures, porosity of the fractures will be highly sensitive to the stress changes. Therefore, we assumed that at the high stress condition of the reservoir, the porosity of the fracture will decrease intensely in comparison to the initial value. To do a sensitivity analysis on this parameter, we considered three values of $a$. Here $c$ is another parameter which controls the porosity dependency on the stress values. Previously, researchers used a value close to 1.5 to include this parameter during the calculation (Rutqvist et al. 2002). Considering this, here a sensitivity analysis is performed by assuming two additional values one higher and one lower value than the base case. Therefore, to consider the sensitivity in both these parameters, nine different scenarios are designed and we investigate the impact of the porosity, stress and permeability relationship on the thermo-hydro-mechanical processes occurring inside the reservoir in Soultz-sous-Forêts (see Table 2).

In each time step, based on the calculated stress field the porosity field will be updated using Eq. [9] and this porosity field will be the basis to calculate the new permeability field for the next step of the simulation. Case 5 from Table 2 is selected as the base case.

Table 2. Properties of the nine scenarios to examine the mechanical impact based on Eq. [9].

| Case number | 1 | 2 | 3 | 4 | 5 | 6 | 7 | 8 | 9 |
|---|---|---|---|---|---|---|---|---|---|
| $c$ | 1 | 1 | 1 | 1.5 | 1.5 | 1.5 | 2 | 2 | 2 |
| $a$ | 0.3 | 0.5 | 0.7 | 0.3 | 0.5 | 0.7 | 0.3 | 0.5 | 0.7 |

Figure 3 shows the temperature, effective stress, porosity and permeability variations of the faulted zone for the one, 30 and 300 years after the operation for the base case. Figure 3 (a1-a3) shows that thermal breakthrough at the bottom hole section is not observed for the entire operation at the given injection and production rates. With the development of cold front around the injection wellbores, thermoelasticity coupled with the poroelasticity decreases the compressive stress (Figure 3(b1-b3), 3(c1-c3) and 3(d1-d3)). Injection flow rate at the surface for the GPK-3 is higher than GPK-4 but most of the fluid does not reach the open hole section due to the high leakage rate. Based on this, we are observing an intense change in temperature and the stress field in the vicinity of GPK-4 rather than GPK-3. Here the results are presented on the surface of the faulted zone which has higher permeability in comparison to the matrix zone. Due to this, the temperature and stress variation for the matrix zone which is not shown here is much smaller. Previously, it has been shown that the poroelasticity effects are concentrated around the wellbores while thermoelasticity can impact on a wider region (Pandey and Vishal 2017; Pandey et al. 2017; Mahmoodpour et al. 2021b, 2022b). Comparing the temperature field with the stress components confirms this behavior and the total stress in the wide region of the reservoir is mainly controlled by the thermoelasticity. Due to the considered free surface boundary condition at the top surface, stress in the vertical direction can release excessive stress in comparison to the roller boundary condition which are considered at the side boundaries. Based on this, horizontal stresses are more in compressive mode. Furthermore, initial stress in the *y*-direction is more in the compressive mode in comparison to the *x*-direction. This results in higher compressive stress in the *y*-direction during the operation period. By injecting cold fluid, stress field increases in tensile mode in all directions which eventuates to increase in the porosity in the effective areas based on the Eq. [9]. Figure 3(e1-e3) shows that porosity increment with respect to the initial value which represents that it may double in the region close to the injection wellbore. It should be noted that porosity variation inside the faulted zone is mainly controlled by the fracture aperture variation which is possible to be doubled (Zhao et al. 2015; Pandey and Vishal 2017; Pandey et al. 2017; Vik et al. 2018; Yao et al. 2018; Salimzadeh et al. 2018; Wang et al. 2019; Yuan et al. 2020; Aliyu et al. 2021; Cui and Wong 2021; Kong et al. 2022). Figure 3(f1-f3) shows the permeability development based on the Eq. [10]. Results represents that a small variation in the porosity may impact the permeability values intensely where in Figure 3(f3), permeability around GPK-4 may increase up to 30 times from the initial value. It is to be noted that in previous studies, Cui et al. (2021) and Salimzadeh et al. (2018) observed the fracture aperture increase up to six and eight times respectively, which is equivalent to 36 and 64 times increase in the fracture permeability. Based on this, this localized permeability increase is well known phenomenon. However, this high permeability changes are mainly restricted in the vicinity the wellbore regions only. It is to be noted that in the reservoir scale calculation, we ignored the fracture initiation and propagation caused by the stress field variations.

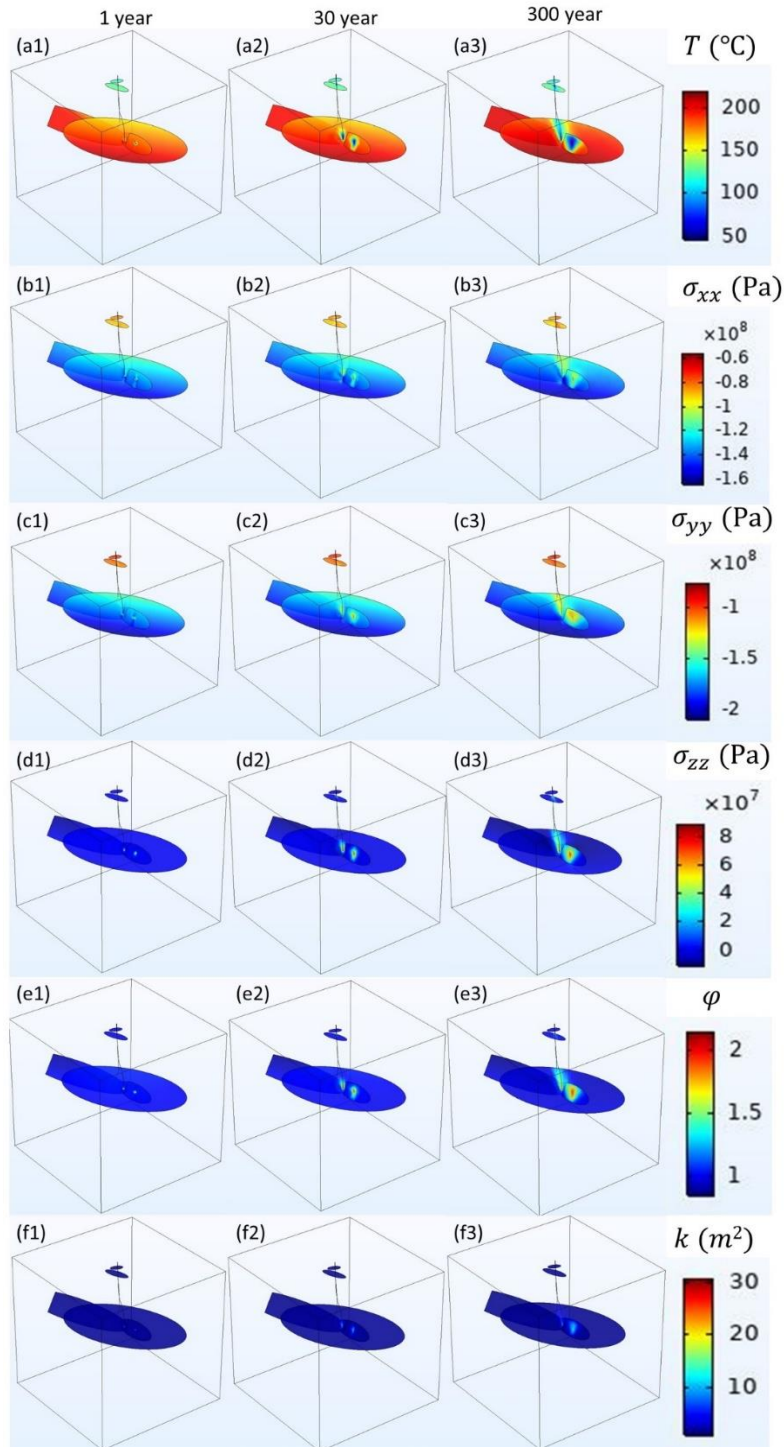

Figure 3: THM results for the base case scenario where *a* is 0.5 and *c* is 1.5. The left column indicates simulation results after one-year, middle column shows results after 30 years and the right column shows results after 300 years for (a1-a3) temperature, (b1-b3) principle stress in *x*-direction, (c1-c3) principle stress in *y*-direction, (d1-d3) principle stress in *z*-direction, (e1-e3) porosity changes ratio compared to the initial values and (f1-f3) permeability changes ratio compared to the initial values.

For better understanding of the THM process governing the fracture aperture changes and heat extraction, the faulted zone of FZ4760 is selected as an example and temperature, porosity and permeability changes for this faulted zone are shown in Figure 4. In the initial time (one year), Figure 4(a1) shows that temperature front is restricted around the GPK-3. Due to the lower compressive stress in the upper section of the faulted zone, porosity and permeability of this area is higher in comparison to the lower section using Eq. [9 & 10]. For the later times (30 years and 300 years), in Figure 4(b1 & c1), the temperature sways towards the top region instead of elliptical temperature distribution between the injection and production wells. Due to the high impact of the thermal stress on the porosity changes, the porosity pattern follows up a similar pattern to the temperature distributions (see Figure 4(a2, b2 & c2)). During the cold fluid injection, pressure increases and synchronously temperature reduction favors the reduction in the compressive stress except in the area closer to the production well. Therefore, in the most part of the reservoir the final to initial porosity value ratio is larger than one. The maximum porosity increment occurs for the region closer to the injection well and its magnitude is approximately 1.6. The permeability field resulting from the porosity changes are shown in Figure 4(a3-c3). The coldest region connecting these two injection-production wellbores shows a permeability increase of approximately six times whereas the previous studies including Yao et al. (2018), Yuan et al. (2020), Wang et al. (2019) and Aliyu et al. (2021) observed fracture permeability increase up to four, five, seven and ten times, respectively. These permeability changes depend on the boundary condition, and fluid and rock properties dependent. The region surrounding the coldest zone has an increase in permeability ratio up to three times and the outermost region shows an increase in permeability ratio by 1.5 times in the doublet region. Beyond the doublet, the porosity and permeability are affected less. Based on these finding, considering a simple horizontal fracture or multiple horizontal fractures in the doublet region provides misleading results in comparison to real case where the fracture dip is not zero. Furthermore, consideration of a mechanical behavior results to a nonsymmetric temperature front between the doublet. Considering the case with fracture initiation and propagation, porosity and permeability ratio increases will be higher due to new fractures. Figure 4(c1-c3) shows that temperature penetration rate is extremely low where after 300 years, just a small area of the reservoir is impacted with the stress field changes. In comparison to the hydraulic fracturing or acidizing process, which are mainly controlled by the pressure pulses and the fluid penetration, thermally induced permeability changes studied in this work are much smaller and slower.

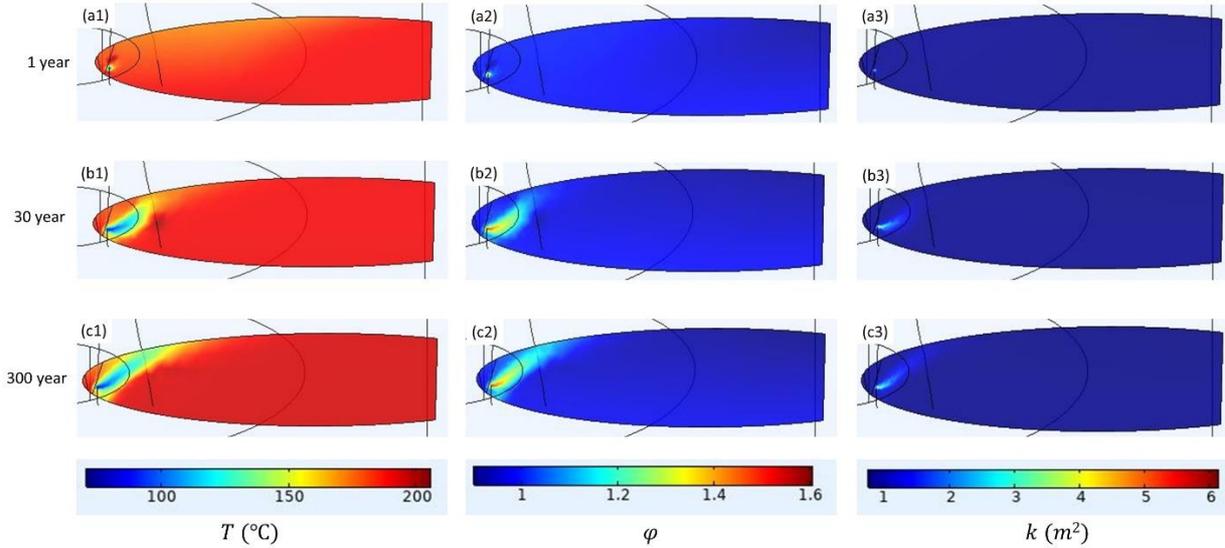

Figure 4: THM results for the fault FZ4760 and the base case stress scenario i.e., *a* is 0.5 and *c* is 1.5. (a1, b1 & c1) temperature (a2, b2 & c2) porosity changes ratio compared to initial values and (a3, b3 & c3) permeability changes compared to initial values. Here the temperature changes inside the fault zone significantly influences the porosity and permeability variation.

To examine the changes in the porosity values in 3D space, isosurface distributions are generated for the base case. Results are shown for the three values of porosity ratio: 1.2 (Figure 5(a1-a3)), 1.5 (Figure 5(b1-b3)), 1.8 (Figure 5(c1-c3)) after the one, 30 and 300 years of operation. It is important to note that there is a leakage along the GPK-2 and GPK-3 and these two wells are much closer to each other in the upper region of the reservoir. This results in higher porosity increase in the upper region and smaller porosity increase at the bottom hole section in the vicinity of GPK-3 at early time (one year). Since there is no leakage for GPK-4, higher porosity increase is observed in the bottom section after one year (see Figure 5(a1)). From Figure 5(a2, a3 & b2, b3) the porosity enhancement inside the faulted zone is higher in comparison to the matrix zone as expected. Due to the higher rate of injection, at the open hole section of GPK-4, wider area of porosity increase zone is observed. Isosurfaces of 1.5 shows slight penetration from the wellbore towards the reservoir while isosurface of 1.8 is visible only for the wellbore region. The porosity increase is restricted to the injection wellbores and there are no detectable changes in the porosity of the bottom hole of GPK-2. Previously, we showed that production rate is a strong function of permeability in the vicinity of the production well (Mahmoodpour et al. 2022a) and the changes around the injection wellbores may not affect the production rate considerably. Based on this, we expect not to see a detectable change in the production flow rate or energy extraction for the entire operational period. Higher conductivity of FZ4925 results in a higher porosity increase inside the faulted zone. However, due to unavailability of any direct connection between this faulted zone and the production wellbore, we cannot expect energy extraction enhancement from this porosity increase. Due to the fluid flow connectivity between GPK-2 and GPK-3 in the upper region, injected cold fluid do not expand much around the injection wellbore for the entire operational duration. The leaked fluid from the injection wellbore reaches to the production wellbore and thus overall fluid loss is limited.

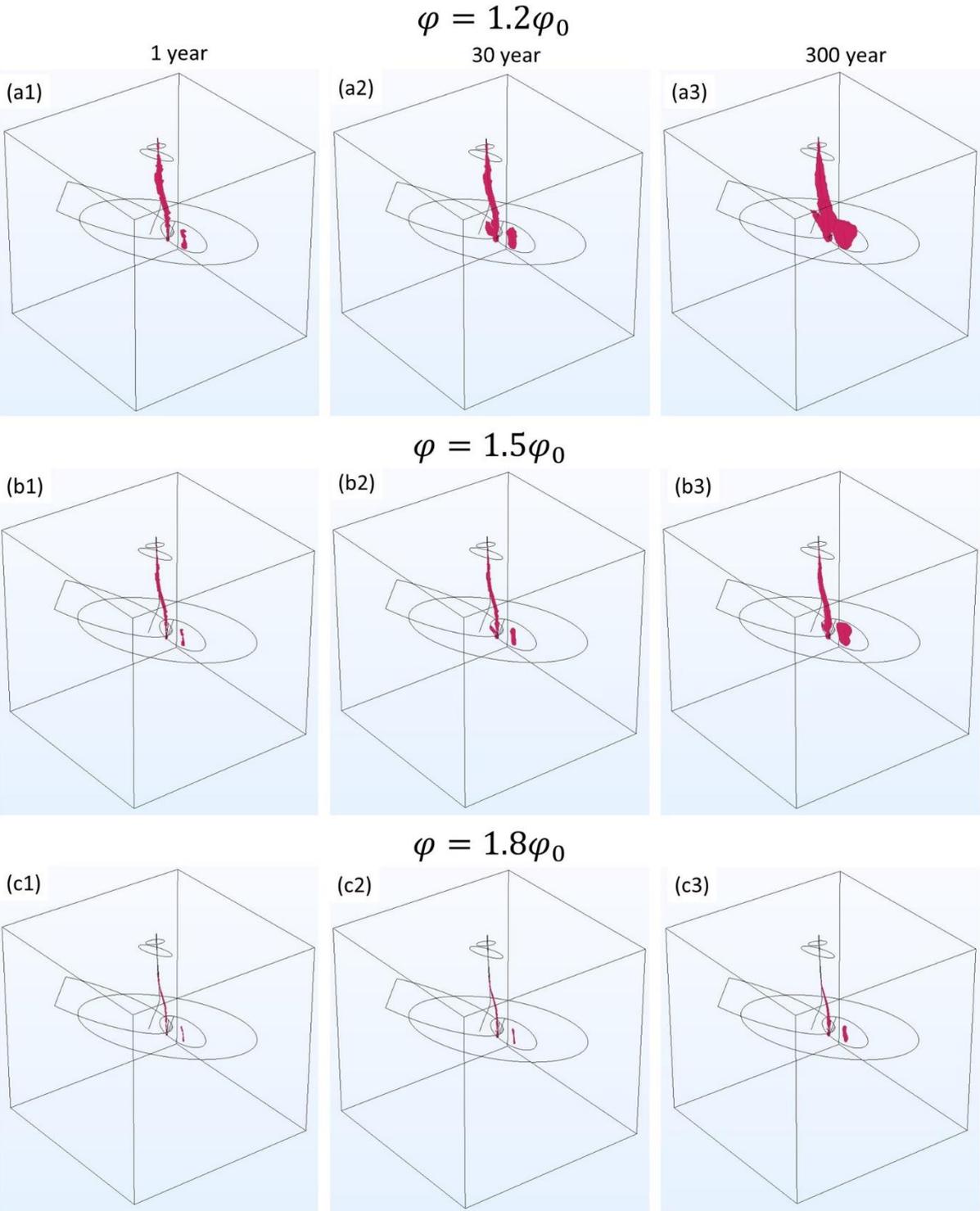

Figure 5: Porosity isosurfaces along the injection wellbores due to THM processes at time one year (left column), 30 years (middle column) and 300 years (right column) for 20% increase from the initial value (a1-a3), 50% increase (b1-b3) and 80 % increase over the initial value (c1-c3).

Further, we have investigated the changes in the permeability of the reservoir and the faulted region values in 3D space using isosurface distributions for the base case. Four different values of permeability increase are shown in the Figure 6: for (a1-a3) 50%. (b1-b3) 200%, (c1-c3) 500% and (d1-d3) 1100% increase from the initial permeability values. Similar to the results for the porosity changes, permeability variation also depicts higher permeability increase in the upper regions of the GPK-2 and GPK-3 at early time however, this enhancement is not sustainable. At later times (30 years and 300 years) it is clearly visible that the permeability increase is much larger in the regions closer to the bottom hole section of the injection wellbores and enhancement inside the FZ4925 is higher in comparison to other faulted zones. The permeability increment near the bottom hole section of the production wellbore is negligible and the resulting production flow rates are smaller. The fluid flowing in GPK-3 and GPK-4 toward GPK-2 remains in contact with the hot rock for a longer period of time and its potential to decrease the temperature of the rock diminishes at the bottom hole region. It is noteworthy that the 500% of permeability increment is restricted to only the wellbores and therefore, it does not affect the production flow rate. However, due to this permeability increment, pressure buildup and the consequent seismic events will decrease. Figure 6(a1) shows that after one year, there is not enough permeability increase in the faulted zone FZ4925 in the vicinity of the production wellbore. However, at time 30 years, there is a large region of high permeability in the FZ4925 towards to the production wellbore. The fluid arriving in this region further increases the permeability of FZ4760 as shown in Figure 6(a3) after 300 years. A similar behavior is clearly evident in Figure 6(b2) and 6(c2). It is interesting to note that permeability does not increases significantly in the leakage region due to a small fluid resident time in that region and it is economically attractive finding.

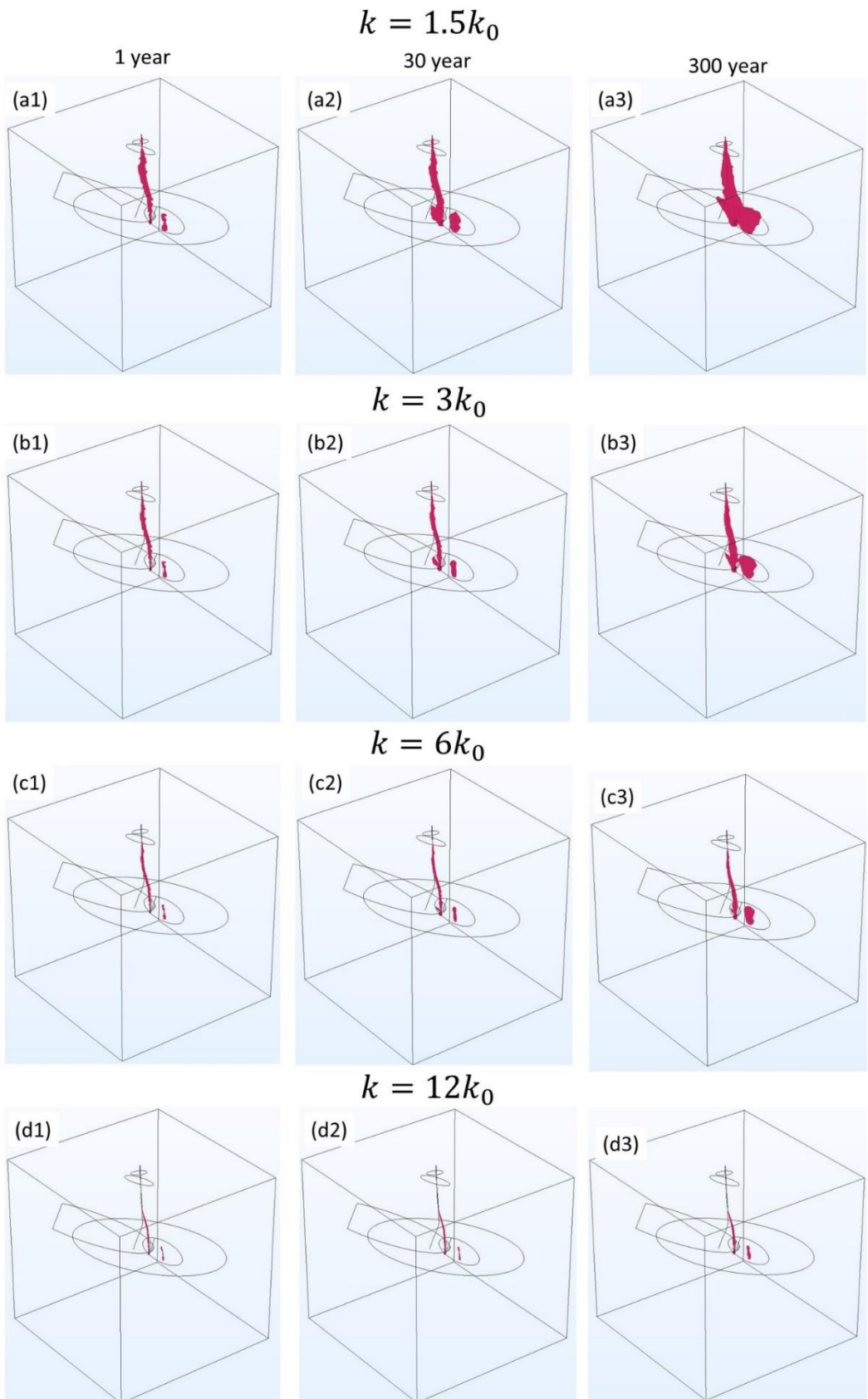

Figure 6: Permeability isosurfaces along the injection wellbores due to THM processes at time one year (left column), 30 years (middle column) and 300 years (right column) for 0.5 times increase from the initial value (a1-a3), 3 times increase (b1-b3), six times increase over the initial value (c1-c3), and 12 times increase over the initial value (c1-c3).

So far, we have presented the THM results for the base case. In the following section, a sensitivity analysis is performed for the parameters governing the geomechanical behavior of the entire reservoir as listed in Table 2. Figure 7 shows permeability isosurfaces where the isosurface values represent 50% increase in permeability with respect to the initial permeability after 300 years of operation. Figure 7(a1, b1 & c1) shows the cases with different values of $c$ and $a = 0.3$. In comparison to the Figure 7(a2, b2 & c2), with $a = 0.5$ and 7(a3, b3 & c3), with $a = 0.7$, the porosity reduction at maximum attainable stress field for cases shown by Figure 7(a1, b1 & c1) demonstrates higher values. Based on this, while the left column has the highest permeability values the right column shows the lowest permeability increase. However, permeability changes are not a strong function of $a$. Here, $c$ is a controlling factor for the permeability changes not only in the bottom hole zone but also in the upper section of the wellbores. Case with $a = 0.3$ and $c = 2$ shows the highest permeability changes between the examined cases. By increasing the $c$ values, high permeability region around the casing leakage zone expands due to the increase in the porosity. In the similar manner, the high permeability region in the bottom holes of the wells for the cases with $c = 2$ is larger. However, after 300 years of the simulation, this region does not contribute in the fluid production near the open hole zone.

From the heat extraction aspect, pressure and temperature changes inside the reservoir are the main controlling factors. To track these changes, the production wellbore is selected and corresponding pressure and temperature variation alongside it is provided in Figure 8 and 9, respectively. Two peaks in the Figure 8 indicates the intersection point of the faulted zones (FZ4760 and FZ4770) and the production wellbore. Figure 8(a) and 8(b) shows the effect of the $a$ on the pressure changes alongside the production well. Comparison reveals that $a$ has a negligible impact on the pressure variation along the GPK-2. Due to the higher conductivity of this faulted zone, pressure loss inside them will be negligible and pressure at the intersecting point will be close to the pressure at the GPK-3 in the same fault. It is to be noted that the faulted zones naming is performed based on the intersection point between GPK-3 and the respective fault. These two faults (FZ4760 and FZ4770) have a closed proximity depth intersection with GPK-3 and their pressure values are close consequently. Based on this, these two peaks in Figure 8 have similar pressure values. As time progresses, permeability inside the matrix zone increases due to thermoporoelastic processes as demonstrated in Figure 6. Also, due to the constant injection and production rates, with higher permeability, pressure gradient between the injection and production well decreases and pressure alongside the production well in the matrix zone increases. However, for the more stress sensitive cases, the presence of the faulted zone may change the increasing trend of the pressure with time (Figure 8(c & d)). For these cases, due to the greater increase in the permeability in the bottom hole section of the production wellbores, the pressure drawdown resulting from the fluid production is higher in comparison to the pressure increases from the fluid injection. Therefore, until the intersecting region between the FZ4760 and the production wellbore, there is a nonmonotonic pressure variation with time.

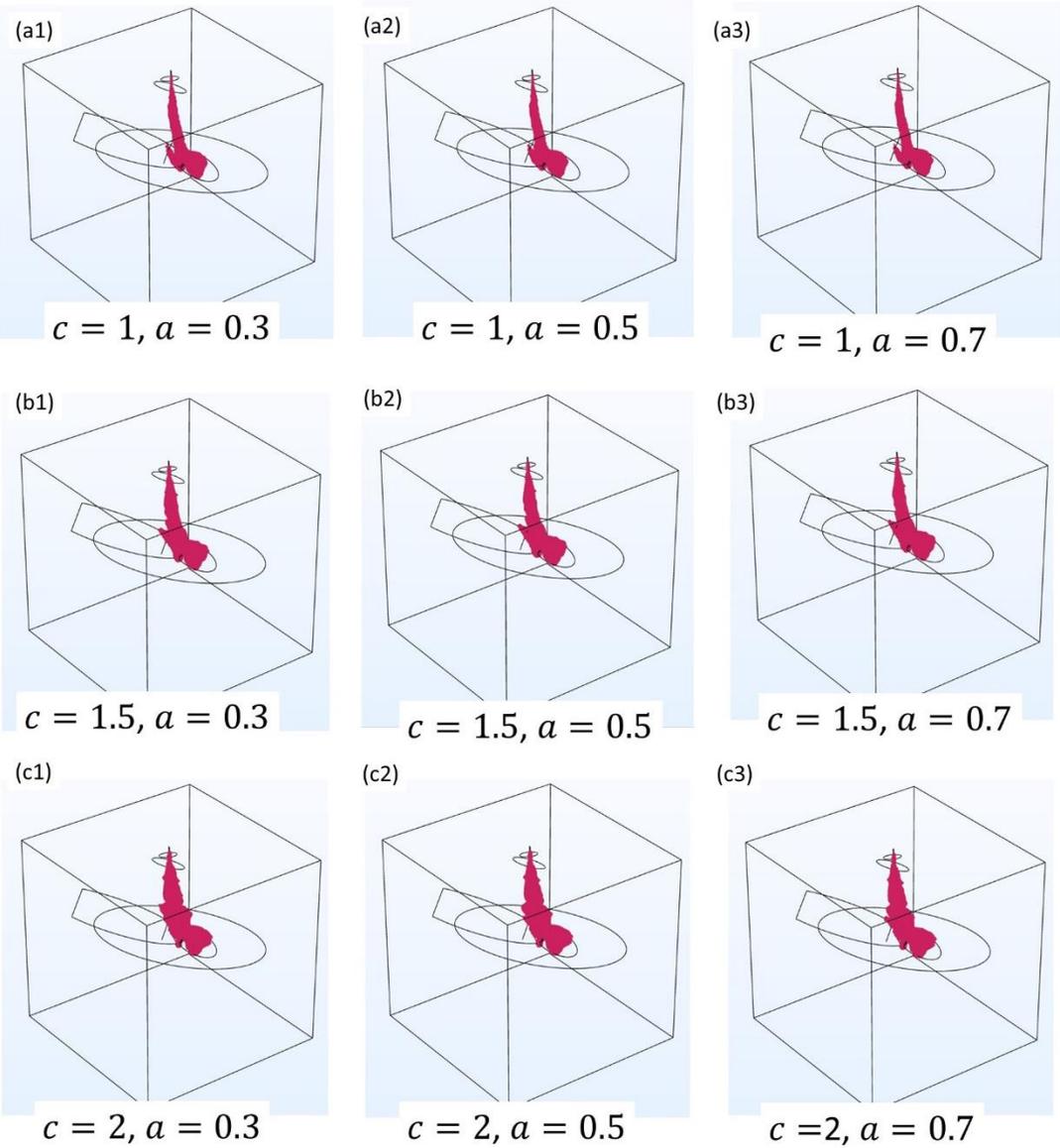

Figure 7: Permeability isosurfaces for 50 % increase over the initial permeability due to THM processes at time 300 years.

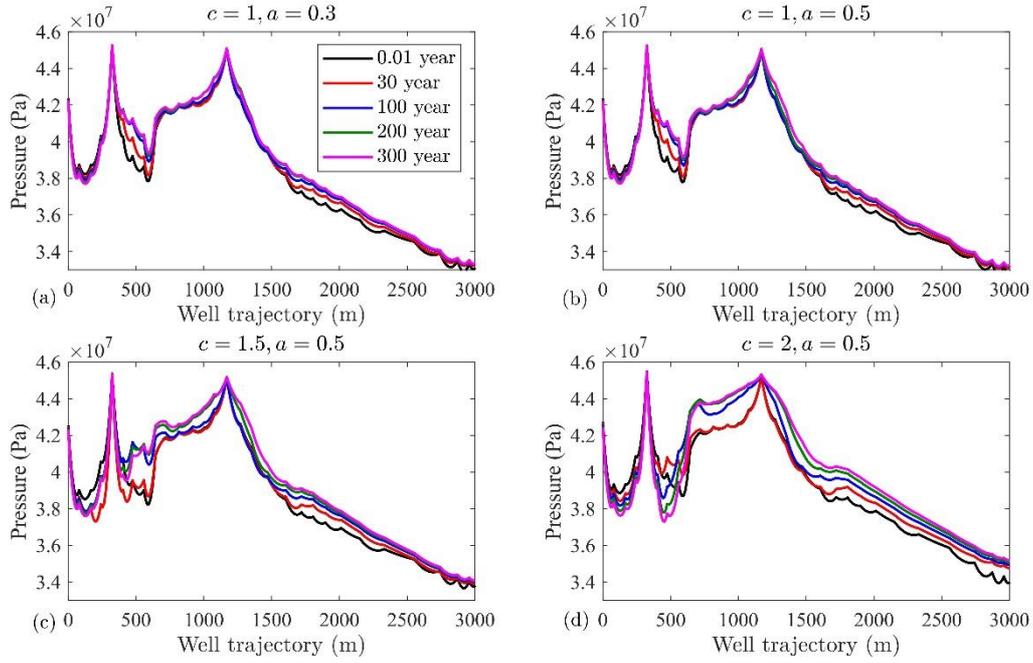

Figure 8: Pressure variation along the production well where the well trajectory is measured from the bottom of the production well.

For the temperature analysis, we start from the base case. Figure 9 (c) shows the temperature variation alongside the GPK-2 for the base case and its sensitivity to the *a* and *c* values (Figure 9(a, b & d)). Figure 9(c) shows a temperature increase in the bottom hole section of GPK-2 in comparison to the initial temperature of the reservoir. This can be explained by Figure 4(left column), where it is clear that the fluid movement is from the bottom toward the top section. Fluid in the bottom zone has a higher temperature and it increases the temperature alongside the GPK-2 near the bottom hole section (red curve in Figure 9(c)). Up to 300 years of operation, the local minimum in the temperature occurs for the intersection of GPK-2 with FZ4770. For this point and the surrounding zone, injected cold fluid impact is quicker due to the higher permeability enhancement. The second local minimum is for the intersection of GPK-2 with the FZ4760. Beyond this region, or at a distance beyond 1500 m from the bottom hole of the GPK-2, there is a rapid temperature reduction caused by the closeness of the GPK-2 and GPK-3. In the open hole region of the GPK-2 (up to a distance almost 1500 m from the bottom), until 100 years, the temperature remains almost constant due to the higher distance between the GPK-2 and GPK-3 except at the intersection of GPK-2 and FZ-4770. However, for 200 years and 300 years, temperature reduction front reaches to the GPK-2 at the intersection point with FZ4760. For the upper regions, temperature reduction front affects drastically GPK-2 in 30 years whereas the change beyond this time is minimal. Figure 9(a & b) are the cases with a little dependency to the stress field changes. Comparing with the Figure 9(c), these cases show negligible temperature increase in the bottom section of GPK-2 due to the lower conductivity of the faulted zone to support the hot fluid. Also, due to the lower support of the FZ4770 in Figure 9(a & b), even 30 years of operation shows a local minimum for the intersection of GPK-3 with the faulted zone. Comparing Figure 9(a & b), it is clear that $a$ has negligible impact on the temperature along the GPK-2. Due to the higher permeability enhancement in the matrix zone of the Figure 9(c), it shows a temperature reduction in the matrix zone between the two faulted zones in comparison to the case shown by Figure 9(b). This confirms that the $c$ value is not only important for the faulted zone but also for the matrix zone. Figure 9(d) shows the intensified picture of this behavior. The temperature alongside GPK-2 decreases with increasing the $c$ value (see Figure 9(b, c & d)).

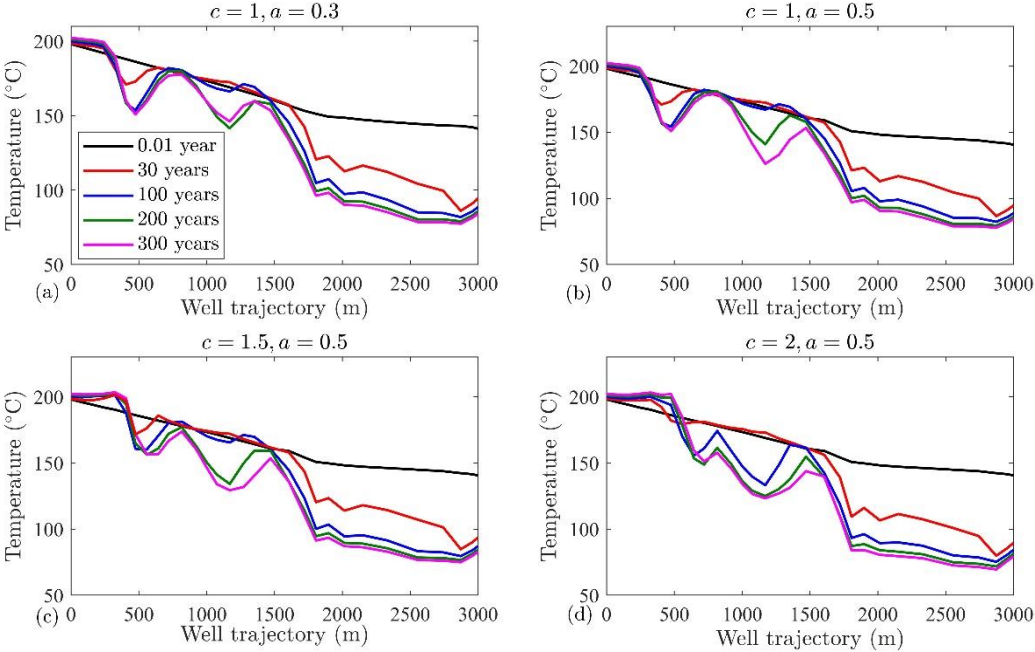

Figure 9: Temperature variation along the production well where the well trajectory is measured from the bottom of the production well.

## 4. Conclusions

A fully coupled thermo-hydro-mechanical process considering thermo-poro-elastic stresses for Soultz-sous-Forêts is numerically modeled. Porosity and permeability alterations in the matrix and the faulted zones are considered. Two mechanical parameters that are necessary to track the coupled nature of the porosity-permeability-stress field is identified. A sensitivity analysis shows that one of them is important and the other is not much essential parameter. The permeability changes are in the range of the reported values for the idealized systems but it does not affect the heat extraction rate at the bottom hole region. Except the faulted zone intersection with the production wellbore, temperature remains almost constant for a long production duration at the bottom hole zone. Fast temperature reduction occurs in the upper zone where there is leakage in the vicinity of injection and production well. Porosity may become double in the vicinity of the injection wellbore but it mainly changes up to 50% of the initial value. In a similar manner, the permeability increases up to 30 times in some regions close to the injection wellbores while the permeability increases up to four times is possible in a wider section surrounding the injection wellbore. In this study, we kept the injection and production rates based on the operational data and performed a sensitivity analysis on the mechanical parameters. In future works, we will consider different operational scenarios (injection and production rates, and temperature) and examined the thermo-poroelastic stress field changes resulting from cold fluid injection. This will build a basis for the seismic event analysis and making a safe EGS operational window.

**Acknowledgement**

The work is conducted as a part of the MEET project that has received funding from the European Union's Horizon 2020 research and innovation program under grant agreement No 792037. Authors have received support from the Group of Geothermal Science and Technology, Institute of Applied Geosciences, Technische Universität Darmstadt. The authors would like to warmly thank the Soultz-sous-Forêts site owner, EEIG Exploitation Minière de la Chaleur, for information access about the Soultz site.